\documentclass{cupconf}
\input epsf

  \checkfont{eurm10}
  \iffontfound
    \IfFileExists{upmath.sty}
      {\typeout{^^JFound AMS Euler Roman fonts on the system,
                   using the 'upmath' package.^^J}%
       \usepackage{upmath}}
      {\typeout{^^JFound AMS Euler Roman fonts on the system, but you
                   dont seem to have the}%
       \typeout{'upmath' package installed. cupconf.cls can take advantage
                 of these fonts,^^Jif you use 'upmath' package.^^J}%
      }
  \else
  \fi

  \checkfont{msam10}
  \iffontfound
    \IfFileExists{amssymb.sty}
      {\typeout{^^JFound AMS Symbol fonts on the system, using the
                'amssymb' package.^^J}%
       \usepackage{amssymb}%
         \let\leq=\leqslant
         
      }{}
  \fi

  \IfFileExists{amsbsy.sty}
    {\typeout{^^JFound the 'amsbsy' package on the system, using it.^^J}%
     \usepackage{amsbsy}}
    {}

\def\hii{H\thinspace{$\scriptstyle{\rm II}$}~}

\def\etal{{\it et al.}~}

\def\eg{{\it e.g.},~}

\def\3he{$^3$He}
\def\4he{$^4$He}
\def\6li{$^6$Li}
\def\7li{$^7$Li}

\def\omegab{$\Omega_{\rm B}$}

\newcommand{\obh}{$\Omega_{\rm B}h^2$}

\def\la{\mathrel{\mathpalette\fun <}}
\def\ga{\mathrel{\mathpalette\fun >}}
\def\fun#1#2{\lower3.6pt\vbox{\baselineskip0pt\lineskip.9pt
  \ialign{$\mathsurround=0pt#1\hfil##\hfil$\crcr#2\crcr\sim\crcr}}}
\def\beq{\begin{equation}}
\def\eeq{\end{equation}}

\newsavebox{\astrutbox}
\sbox{\astrutbox}{\rule[-5pt]{0pt}{20pt}}

\title[Tracking The Baryon Density]{Tracking The Baryon Density
From The Big Bang To The Present}

\author[Gary Steigman]
{G\ls A\ls R\ls Y\ns  S\ls T\ls E\ls I\ls G\ls M\ls A\ls N}

\affiliation{Departments of Physics and Astronomy, The Ohio State 
University, Columbus, OH 43210, USA\\}

\begin{document}

\maketitle

\begin{abstract}
The primordial abundances of deuterium, helium, and lithium probe the 
baryon density of the universe only a few minutes after the Big Bang.  
Of these relics from the early universe, deuterium is the baryometer 
of choice.  After reviewing the current observational status (a moving 
target!), the BBN baryon density is derived and compared to independent 
estimates of the baryon density several hundred thousand years after 
the Big Bang (as inferred from CMB observations) and at present, more 
than 10 billion years later.  The excellent agreement among these values 
represents an impressive confirmation of the standard model of cosmology, 
justifying -- indeed, demanding -- more detailed quantitative scrutiny.  
To this end, the corresponding BBN-predicted abundances of helium and 
lithium are compared with observations to further test and constrain 
the standard, hot, big bang cosmological model.
\end{abstract}

\firstsection 
\section{Introduction}

As progress is made towards a new, precision era of cosmology, 
{\it redundancy} will play an increasingly important role.  As 
cosmology is an {\it observational} science, it will be crucial 
to avail ourselves of multiple, independent tests of, and constraints 
on, competing cosmological models and their parameters.  Furthermore, 
such redundancy may provide the only window on systematic errors
which can impede our progress or send us off in unprofitable
directions.  To illustrate the efficacy of such an approach in 
modern cosmology, I'll track the baryon density of the Universe as 
revealed early on (first few minutes) by Big Bang Nucleosynthesis 
(BBN), later (few hundred thousand years) as coded in the fluctuation 
spectrum of the Cosmic Microwave Background (CMB) radiation, and 
up to the present, approximately 10 Gyr after the expansion began.  
As theory suggests and terrestrial experiments confirm, baryon number 
should be preserved throughout these epochs in the evolution of the 
universe, so that the number of baryons ($\equiv$~nucleons) in a 
comoving volume {\it should} be unchanged from BBN to today.  As 
a surrogate for identifying a comoving volume, we may compare the 
baryon/nucleon density to the density of CMB relic photons.  Except 
for the additional photons produced when e$^{\pm}$ pairs annihilate, 
the number of photons in our comoving volume is also preserved.  As 
a result, the baryon density may be tracked through the evolution 
of the universe utilizing the nucleon-to-photon ratio $\eta$ where,
\beq
\eta \equiv n_{\rm N}/n_{\gamma}.
\eeq 
Since the temperature of the CMB fixes the present number density
of relic photons, the fraction of the critical mass/energy density
in baryons today ($\Omega_{\rm B} \equiv \rho_{\rm B}/\rho_{crit}$)
is directly related to $\eta$,
\beq
\eta_{10} \equiv 10^{10}\eta = 274\Omega_{\rm B}h^{2},
\eeq
where the Hubble parameter, measuring the present universal expansion
rate, is H$_{0} \equiv 100h$~kms$^{-1}$Mpc$^{-1}$.  According to the 
HST Key Project, $h = 0.72 \pm 0.08$ (\cite[Freedman \etal 2001]{HST}).

For several decades now the best constraints on $\eta$ have come from
the comparison of the predictions of BBN with the observational data
relevant to inferring the primordial abundances of the relic nuclides
D, \3he, \4he, and \7li.  For recent reviews, the interested reader 
is referred to \cite{OSW}, \cite{BNT}, \cite{cyburt}, and references
therein.  However, the time is rapidly approaching when new data of 
similar accuracy will be available to probe the baryon density at 
later epochs.  Indeed, recent CMB data has very nearly achieved
this goal.  This same level of precision is currently lacking for the 
present universe estimates but, such as they are, they do permit us to 
compare and contrast independent estimates of $\eta$ (or \obh) at three 
widely separated eras in the evolution of the universe.  In the next 
section the BBN bounds on the nucleon density are presented and their 
implications for the dark baryon and dark matter problems described.  
Next, we turn to the CMB estimates of the baryon density, comparing 
them to those from BBN.  Last (but not least), we turn to an estimate 
of the present universe baryon density utilizing data whose interpretation 
avoids the need to adopt a relation between mass and light in the universe.

In recent years cosmological research has been data-driven and we have
become used to looking forward with great anticipation to the latest
observational results, and greedily wishing for more.  Be careful of 
what you wish for!  Not all the new data has led us unerringly along 
the right path.  Indeed, within a few weeks of this meeting new data 
became available on the CMB angular fluctuation spectrum (Halverson 
\etal 2001; Netterfield \etal 2001; Lee \etal 2001) and on the deuterium 
abundance (\cite[Levshakov, Dessauges-Zavadsky, D'Odorico, \& Molaro 
2001]{DOd2}; \cite[Pettini \& Bowen 2001]{PB}) which change dramatically 
some of the results I presented in my talk at this Symposium.  In the 
interest of preserving some of the historical record, I will comment 
on the older data and their implications, while reserving the most 
recent observations for my conclusions regarding the baryon density 
of the universe.

\section{The Baryon Density During The First Few Minutes}\label{sec:BBN}

\begin{figure}
  \centering
  \epsfysize=3.5truein
  \epsfbox{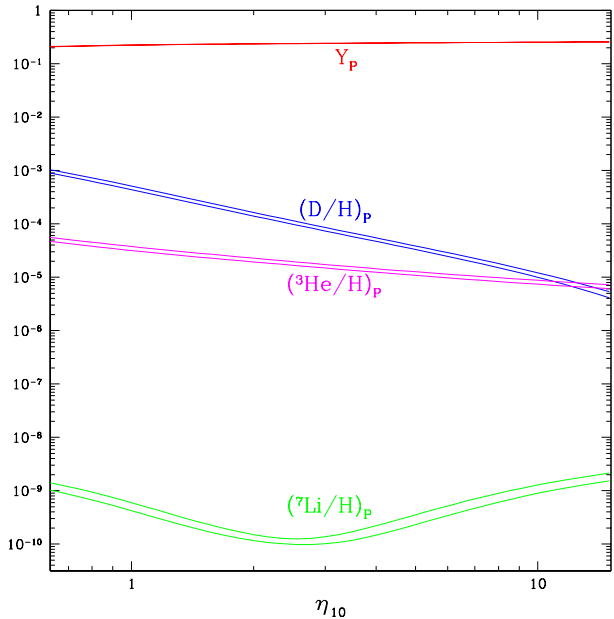}
  \caption{The SBBN-predicted abundances of D, \3he, \7li (by number 
relative to hydrogen) and the \4he mass fraction Y, as a function 
of the nucleon-to-photon ratio $\eta$.  The widths of the bands 
reflect the BBN uncertainties associated with the nuclear and weak 
interaction rates.}\label{fig:bbn}
\end{figure}

During the first few minutes in the evolution of the universe the density
and temperature were sufficiently high for nuclear reactions to occur 
in the time available.  As the universe expanded, cooled, and became more 
dilute, the ``rolling blackout" became permanent and the universal nuclear 
reactor went offline.  The abundances of the light nuclei formed during
this limited epoch of primordial alchemy were determined, for the most 
part, by the competition between the time available (the universal expansion 
rate) and the density of the reactants: nucleons (neutrons and protons).  
The abundances predicted for nucleosynthesis in the standard cosmological 
model (SBBN) are shown (as a function of $\eta$) in Figure~\ref{fig:bbn}.

The abundances of D, \3he, and \7li are ``rate limited", being determined
by the competition between the nuclear production/destruction rates and
the universal expansion rate.  As such, they are sensitive to the nucleon
density and have the potential to serve as ``baryometers".  In contrast,
the nuclear reactions which build \4he are so rapid and, because its 
destruction is inhibited (due to its strong binding and the lack of a 
stable nuclide at mass-5), the helium-4 mass fraction, Y, is insensitive 
to $\eta$.  Since virtually all the neutrons available at BBN are 
incorporated into \4he, Y is determined mainly by the neutron-to-proton 
ratio which is controlled by the competition between the weak interaction 
rate (``$\beta$-decay") and the expansion rate.  As a result, since 
the neutron lifetime is known very accurately, the \4he abundance is 
a sensitive probe of the early universe expansion rate.

\subsection{Deuterium -- The Baryometer Of Choice}

Of the three relic nuclides whose primordial abundances may be probes
of the baryon density (D, \3he, \7li), deuterium is the baryometer of
choice.  First and foremost, the predicted primordial abundance has a
sigificant dependence on the nucleon density (D/H $\propto \eta^{-1.6}$).
As a result, if the primordial abundance is known to, say, 10\%, then
the baryon density ($\eta$) can be determined to $\sim 6\%$; truly
precision cosmology!  Equally important, as \cite{els} showed long ago, 
BBN is the only astrophysical site where an ``interesting" abundance of 
deuterium may be produced (D/H $\ga 10^{-5}$); the relic abundance is 
not enhanced by post-BBN production.  Furthermore, as primordial gas is 
cycled through stars, deuterium is completely destroyed (because of the 
small binding energy of the deuteron, destruction occurs during the 
pre-main sequence evolution, when the stars are fully mixed).  As a 
result, the abundance of deuterium has only decreased (or, remained 
close to its primordial value) since BBN.
\beq
(D/H)_{\rm NOW} \leq (D/H)_{\rm BBN},
\label{d/h1}
\eeq
where ``NOW" refers to the ``true" deuterium abundance in any system
observed at any time (\eg the solar system; the interstellar medium;
high redshift, low metallicity QSO absorbers, etc.).

There is, however, one potentially serious problem associated with 
using deuterium as a baryometer.  The atomic spectra of deuterium 
and hydrogen are identical, save for the isotope shift associated with
the deuteron-proton mass difference.  As a result, a small amount
of hydrogen at the ``wrong" velocity (an ``interloper") can masquerade
as deuterium, so that
\beq
(D/H)_{\rm NOW} \leq (D/H)_{\rm OBS}.
\label{d/h2}
\eeq
A comparison of equations \ref{d/h1} and \ref{d/h2} reveals the problem: 
how to relate
(D/H)$_{\rm OBS}$ to (D/H)$_{\rm BBN}$?  One approach is to concentrate
on observing deuterium in those high redshift (high-z), low metallicity
(low-Z) systems where not much gas will have been cycled through stars
and the deuterium (``NOW") should be essentially primordial (BBN).  If
so, there should be no variation (outside of the statistical errors) with 
metallicity.  To test for interlopers the best approach (the favorite of 
theorists!) is to have lots of data.  Observers are making great progress 
towards this goal, but the road has not always been straight.

After some false starts, there were four high-z, low-Z QSO absorption line
systems where deuterium had been detected as of the time of this Symposium
in April 2001 (Burles \& Tytler 1998a,b; O'Meara \etal 2001; \cite[D'Odorico, 
Dessauges-Zavadsky, \& Molaro 2001]{DOd1}).  A fifth system (\cite[Webb 
\etal 1997]{hid}), with possibly a much higher D/H, is widely agreed to 
have insufficient velocity data to rule out contamination by a hydrogen 
interloper (\cite[Kirkman \etal 2001]{antihid}).  However, these data on 
D/H, displayed as a function of metallicity in Figure 2, appear to challenge 
our expectations (of a low-metallicity, deuterium ``plateau") in that they 
suggest an anticorrelation between D/H and metallicity.  Has deuterium 
actually been destroyed since BBN, and is the BBN abundance at least as 
high as the highest data point?  This would be most surprising in such 
low metallicity systems (see, \eg Jedamzik \& Fuller 1997).
\begin{figure}
  \centering
  \epsfysize=3.5truein
  \epsfbox{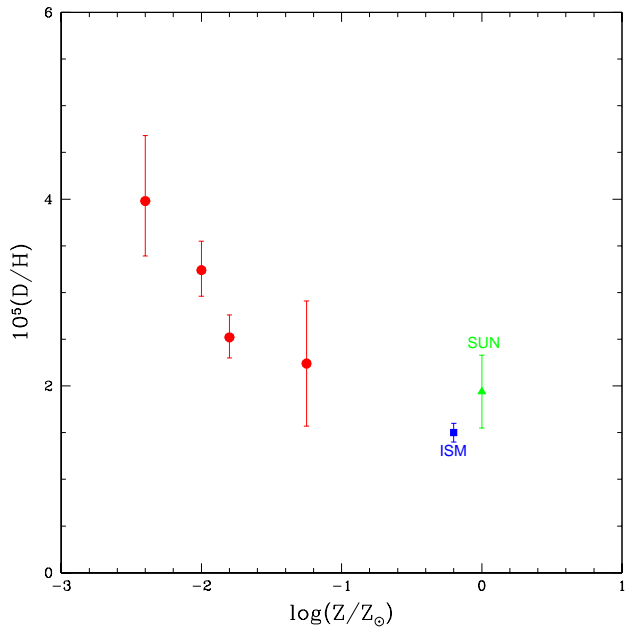}
  \caption{The deuterium abundance by number relative to hydrogen derived 
from observations of high-z, low-Z QSO absorption line systems as a function 
of metallicity (relative to solar).  Also shown are the deuterium abundances 
inferred for the local insterstellar medium (Linsky \& Wood 2000) and for 
the presolar nebula (Gloeckler \& Geiss 2000).}\label{fig:dvsz1}
\end{figure}
Perhaps one or more of the ``high" points is ``too high" because of an
interloper.  To test for this possibility, let's display the same data
as a function of the hydrogen column density (O'Meara \etal 2001); see 
Figure 3.  It might be expected that the lower hydrogen column density
absorbers could be more easily contaminated.  Two of the systems have
relatively low H-column densities ($\la 10^{18}$cm$^{-3} \equiv$ Lyman
Limit Systems (LLS)) and two have considerably higher N$_{\rm H}$ ($\ga
10^{19}$cm$^{-3} \equiv$ Damped Lyman Alpha (DLA) absorbers).
\begin{figure}
  \centering
  \epsfysize=3.5truein
  \epsfbox{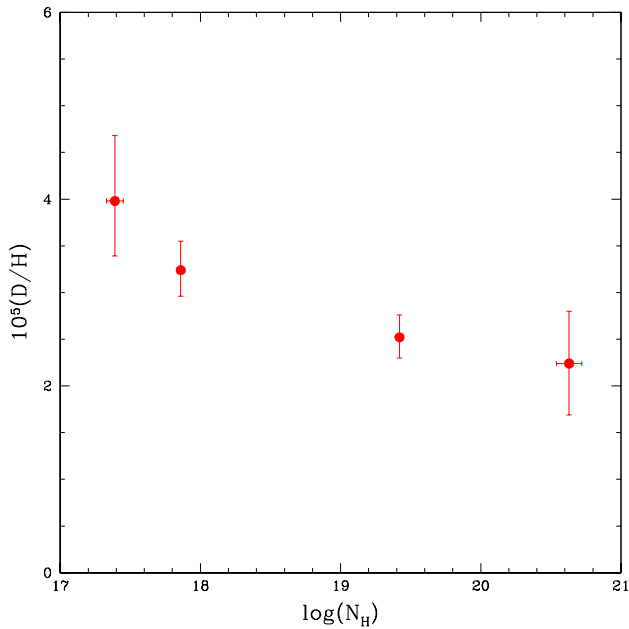}
  \caption{The same deuterium abundances as in Figure 2 plotted as a 
function of the hydrogen column density N$_{\rm H}$ (cm$^{-3}$) in the 
absorbing cloud.}\label{fig:dvsh1}
\end{figure}

Indeed, the hint from Figure 3 is that interlopers may be playing a role
in either or both of the LLS.  If so, perhaps we should identify only the
DLAs with the primordial deuterium abundance.  However, given the small
number of data points, this would be premature.  At the Symposium I adopted
a weighted average ($\langle$D/H$\rangle = 2.9 \pm 0.3 \times 10^{-5}$)
and called for more data.  Observers weren't long in responding.  But, 
I hadn't anticipated what they would find.

Further observations by \cite{DOd2} of the \cite{DOd1} absorber (the 
low D/H point at the highest N$_{\rm H}$ in Fig.~3) revealed a more 
complex velocity structure and led to an upward revision in the derived 
D/H (which, by the way, is by more than the previous statistical error 
estimate, reinforcing the potential for systematic errors to wreak havoc).  
So far, so good, now that the four D/H determinations are more consistent
with one another, alleviating the need to invoke interloper contamination.
However, in the meanwhile \cite{PB} weighed in with a new deuterium
detection in a DLA.  The Pettini-Bowen abundance is smaller than all
the other determinations, indeed smaller than (although within the 
errors of) the presolar nebula abundance of \cite{gg}.  The current 
data (as of July 2001) are shown in Figure 4.
\begin{figure}
  \centering
  \epsfysize=3.5truein
  \epsfbox{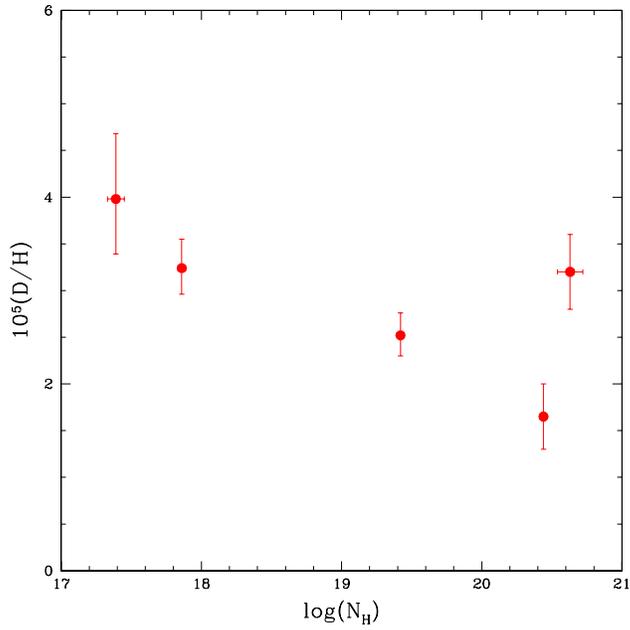}
  \caption{The current (July 2001) deuterium abundances plotted as a 
function of the hydrogen column density N$_{\rm H}$ (cm$^{-3}$) in the 
absorbing cloud.  See the text for references.}\label{fig:dvsh2}
\end{figure}

What to do?  The dispersion among these four determinations hints that
one or more may be wrong.  Until this puzzle is resolved by more data
(the last refuge of the theorist), I will adopt as a default estimate
the one derived by \cite{O'M}: (D/H)$_{\rm BBN} \equiv 3.0 \pm 0.4 
\times 10 ^{-5}$.  The consequences for the BBN derived baryon density
are shown in Figure 5 where the overlap between the predicted and
``observed" primordial abundances are used to constrain $\eta$.     
\begin{figure}
  \centering
  \epsfysize=3.5truein
  \epsfbox{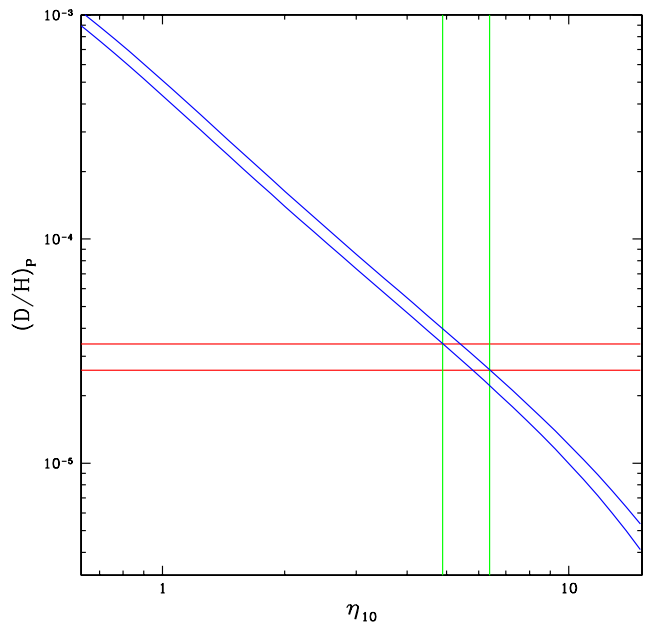}
  \caption{The band stretching from upper left to lower right is the
BBN-predicted deuterium abundance (as in Fig.~1).  The horizontal band 
is the observational estimate of the primordial abundance (see the 
text).  The vertical band provides an estimate of the BBN-derived 
baryon density.}\label{fig:dvseta}
\end{figure}

\subsection{The BBN Baryon Density}

From a careful comparison between the BBN predicted abundance (including
errors -- which are subdominant) and the adopted primordial value, the
baryon density when the universe is less than a half hour old may be
constrained (see Figure 6).  For cosmology, this is truly a ``precision"
determination.  Whether it is accurate, only time will tell.
\beq
\eta_{10}({\rm BBN}) = 5.6 \pm 0.5 ~~~ (\Omega_{\rm B}h^{2} = 0.020 \pm 0.002)
\label{etabbn}
\eeq

\begin{figure}
  \centering
  \epsfysize=3.0truein
  \epsfbox{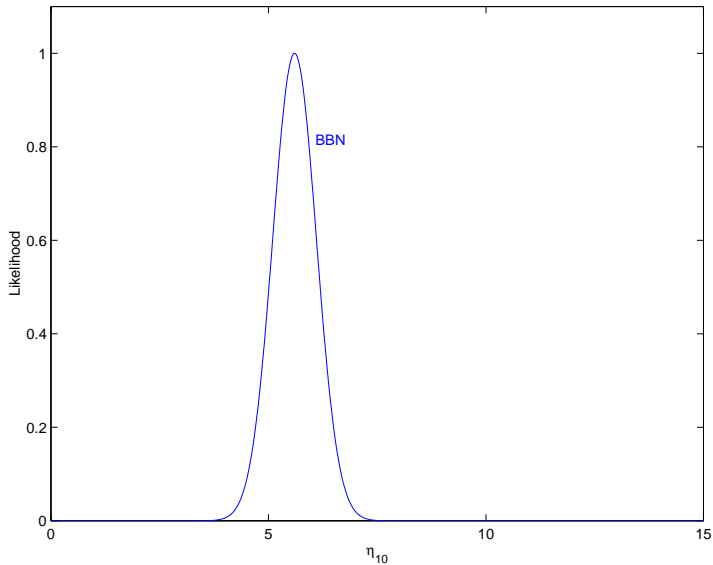}
  \caption{The likelihood distribution (normalized to unity at maximum) 
for the baryon-to-photon ratio derived from BBN and the (adopted) primordial 
abundance of deuterium.}\label{fig:lik1}
\end{figure}

\begin{figure}
  \centering
  \epsfysize=3.5truein
  \epsfbox{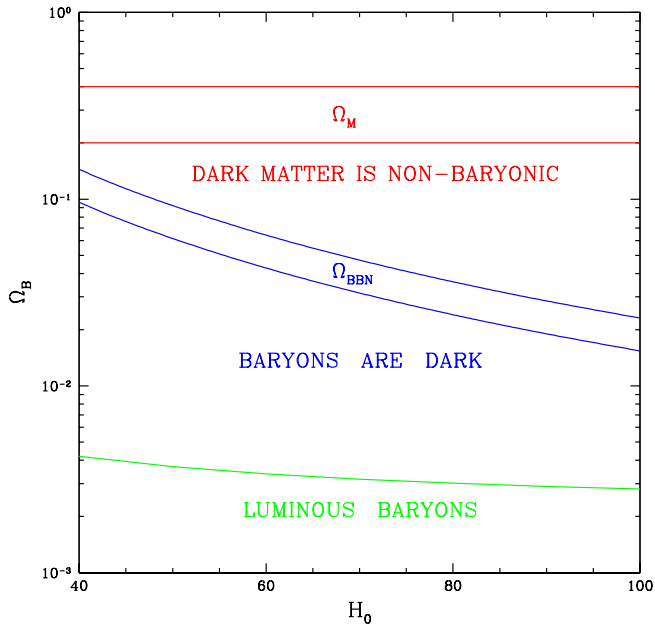}
  \caption{Several estimates of the baryon density (relative to the critical
density) in the present universe as a function of the Hubble parameter
H$_{0}$.  The band labelled ``BBN" is the early universe baryon density
estimate (see eq.~\ref{etabbn}).  The line labelled ``Luminous Baryons" 
is the upper bound to estimates of the density of baryons seen at 
present in emission or absorption (Persic \& Salucci 1992).  The band 
labelled $\Omega_{\rm M}$ ($= 0.3 \pm 0.1$) is an estimate of the total
mass density in nonrelativistic particles at present.}\label{fig:obvsh}
\end{figure}

This range for the baryon density poses some interesting challenges to 
our view of the universe.  These challenges may be seen by comparing the 
BBN estimate of \omegab ~with those found by adding up all the baryons 
associated with the ``luminous" material observed in the present/recent 
universe at z $\la 1$ (Persic \& Salucci 1992), and with estimates of 
the total mass density at present.  These comparisons are shown in 
Figure~\ref{fig:obvsh} where the various density parameter estimates/ranges 
are plotted as a function of the Hubble parameter H$_{0}$ (recall that, 
according to \cite{HST}, H$_{0} = 72 \pm 8$).

The gap between the upper limit to luminous baryons and the BBN band is 
the ``dark baryon problem": not all the baryons expected from BBN have 
been seen in the present universe.  Perhaps it is hubris to expect that 
all baryons will choose to radiate (or absorb) in those parts of the
spectrum we can see, or which our instruments can record.  Indeed, from 
the absorption observed in the Lyman-alpha forest at redshifts of order
2 -- 3 (\cite[Weinberg, Miralda-Escud\'e, Hernquist, \& Katz 1997]{wein}), 
it seems clear that the density of baryons in the present universe is much 
larger than the upper bound to luminous baryons shown in Figure~\ref{fig:obvsh}.  
We return to a different estimate of the baryon density in the present/recent 
universe in \S\,\ref{sec:sn1a}.

The gap between the BBN band and the band labelled $\Omega_{\rm M}$ is
one aspect of the classical dark matter problem: the mass density inferred
from the motions of galaxies exceeds the baryon density derived from BBN.
Aside from those solutions which resort to modifying gravity (\eg Sanders
(2001) this volume), the standard assumption is that this gap provides 
evidence for the dominance at present of non-baryonic dark matter.  Much 
of this Symposium was devoted to discussions of possible dark matter
candidates and the means for their detection. 

\section{The Baryon Density At A Few Hundred Thousand Years}\label{sec:cmb}

\begin{figure}
  \centering
  \epsfysize=3.83truein
  \epsfbox{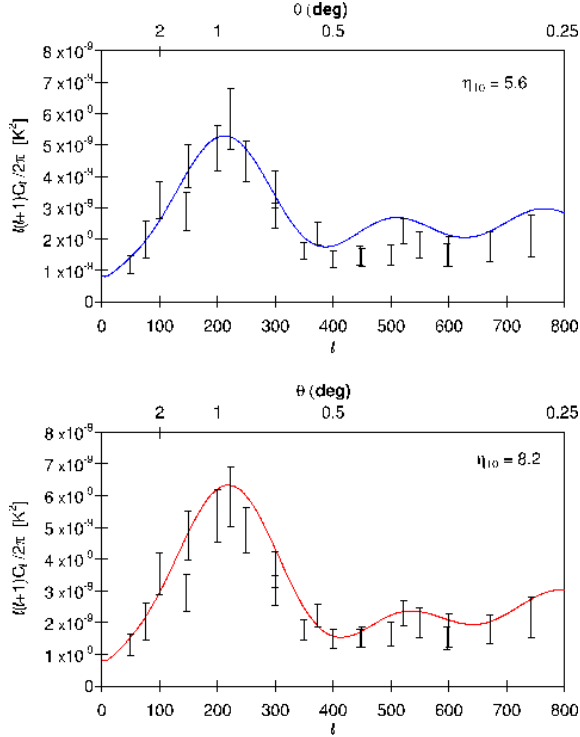}
  \caption{The CMB angular fluctuation spectra for two models which 
differ only in the adopted baryon density.  The BBN inferred baryon 
density is shown in the upper panel and, for comparison, a higher 
baryon density model is shown in the lower panel.  The data are 
from the ``old" BOOMERANG and MAXIMA-1 observations; see the text 
for references.}\label{fig:cmb}
\end{figure}

The early universe is radiation dominated.  As the universe expands and 
cools, the density in non-relativistic matter becomes relatively more 
important, eventually dominating after a few hundred thousand years.  At 
this stage perturbations can begin to grow under the influence of gravity 
and, on scales determined by the relative density of baryons, oscillations 
in the baryon-photon fluid will develop.  When, at a redshift $\sim 1100$, 
the electron-proton plasma combines to form neutral hydrogen, the CMB photons
are freed to propagate thoughout the universe.  These CMB photons preserve
the record of the baryon-photon oscillations through very small temperature
fluctuations in the CMB spectrum which have been detected by the newest
generation of experiments, beginning with COBE (\cite[Bennett \etal 1996]
{cobe}) and continuing with the exciting BOOMERANG (\cite[de Bernardis 
\etal 2000]{boom1}; \cite[Lange \etal 2000]{boom2}) and MAXIMA-1 
(\cite[Hanany \etal 2000]{max1}) results which appeared somewhat more 
than a year ago.  In Figure~\ref{fig:cmb} we illustrate the status quo 
ante with the dramatic and challenging results from those earlier data.

The relative heights of the odd and even peaks in the CMB angular fluctuation 
spectrum depend on the baryon density and the early BOOMERANG and MAXIMA-1 
data favored a ``high" baryon density (compare the ``BBN case", $\eta_{10} 
= 5.6$, in the upper panel with that for a baryon density some 50\% higher 
shown in the lower panel).  At the time of this Symposium, these were the 
extant data and they posed a challenge to the consistency of the standard 
model of cosmology.  

At the Symposium we were told that new data would be forthcoming shortly and, 
the observers didn't disappoint.  In less than a month new (and some revised)
data appeared (Halverson \etal 2001; Netterfield \etal 2001; Lee \etal 2001)
which have eliminated the challenge posed by the older data.  Although the
extraction of cosmological parameters from the CMB data can be very dependent
on the priors adopted in the analyses (see \cite[Kneller \etal 2001]{kssw}),
the inferred baryon density is robust.  \cite{kssw} find,
\beq
\eta_{10}({\rm CMB}) = 6.0 \pm 0.6  ~~~ (\Omega_{\rm B}h^{2} = 0.022 \pm 0.002).
\eeq

\begin{figure}
  \centering
  \epsfysize=3.0truein
  \epsfbox{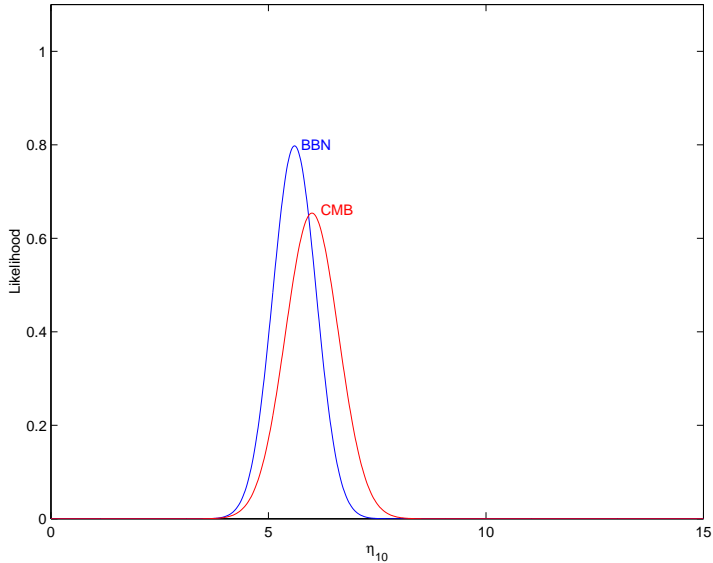}
  \caption{The likelihood distributions (normalized to equal areas under 
the curves) for the baryon-to-photon ratios derived from BBN and from the 
CMB.}\label{fig:lik2}
\end{figure}

In Figure~\ref{fig:lik2} is shown the comparison of the likelihoods for
the nucleon-to-photon ratio from BBN, when the universe was some tens of
minutes old, and from the CMB, some few hundred thousand years later.
The excellent agreement between the two, independent estimates is a
spectacular success for the standard model of cosmology and an illustration
of the great potential for precision tests of cosmology.

\section{The Baryon Density At 10 Gyr}\label{sec:sn1a}

As already discussed in \S\,\ref{sec:BBN} (see Figure~\ref{fig:obvsh}), 
the amount of baryons ``visible" in the present universe is small compared 
to that expected on the basis of BBN (and, as seen in \S\,\ref{sec:cmb}, 
to that revealed by the CMB data).  If most baryons in the present universe 
are dark, how can their density be constrained?  There are a variety of 
approaches.  Many depend on assumptions concerning the relation between 
mass and light, or require the adoption of a specific model for the growth 
of structure.  In the approach utilized here we attempt to avoid such 
model-dependent assumptions.  Instead, we use the data from the SNIa 
magnitude-redshift surveys (\cite[Perlmutter \etal 1997]{scp1}; \cite[Schmidt 
\etal 1998]{hiz}; \cite[Perlmutter \etal 1999]{scp2}), along with the 
{\it assumption} of a flat universe (which receives strong support from 
the newest CMB data (Halverson \etal 2001; Netterfield \etal 2001; Lee 
\etal 2001)) to pin down the total matter density ($\Omega_{\rm M}$),
which will be combined with an estimate of the universal baryon
fraction ($f_{\rm B} \equiv \Omega_{\rm B}/\Omega_{\rm M}$) derived
from studying the X-ray emission from clusters of galaxies.  For
more details on this approach, see \cite{shf} and \cite{swz}.  

\begin{figure}
  \centering
  \epsfysize=3.0truein
  \epsfbox{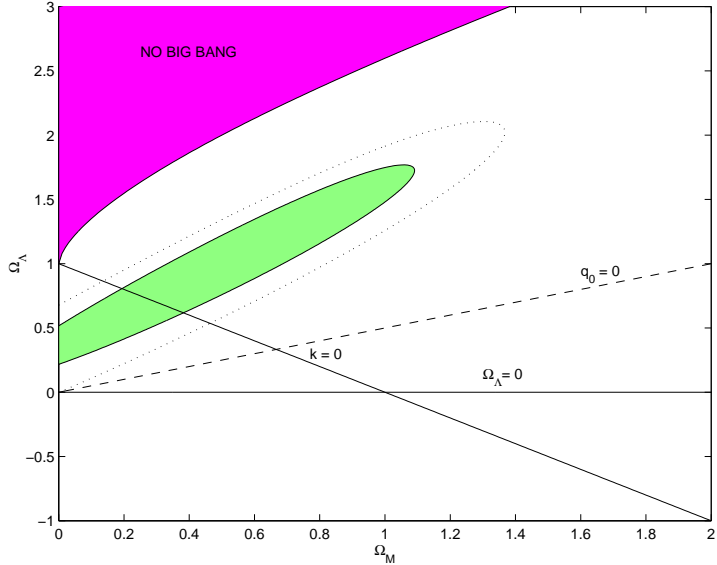}
  \caption{The 68\% (solid) and 95\% (dotted) contours in the 
$\Omega_{\Lambda} - \Omega_{\rm M}$ plane allowed by the SNIa
magnitude-redshift data (see the text for references).  
Geometrically flat models lie along the diagonal line labelled 
$k = 0$.}\label{fig:sn1a}
\end{figure}

In Figure~\ref{fig:sn1a} are shown the SNIa-constrained 68\% and 
95\% contours in the $\Omega_{\Lambda} - \Omega_{\rm M}$ plane.
The expansion of the universe is currently accelerating for those
models which lie above the (dashed) q$_{0} = 0$ line.  The $k = 0$
line is for a ``flat" (zero 3-space curvature) universe.  As shown 
in \cite{swz}, adopting the assumption of flatness and assuming the 
validity of the SNIa data, leads to a reasonably accurate ($\sim 25\%$)
estimate of the present matter density.
\beq
\Omega_{\rm M}(\rm{SNIa;Flat}) = 0.28 ^{+0.08} _{-0.07}.
\eeq

As the largest collapsed objects, rich clusters of galaxies
provide an ideal probe of the baryon {\it fraction} in the
present universe.  Observations of the X-ray emission from 
clusters of galaxies permit constraints on the hot gas
content of such clusters which, when corrected for baryons
in stars (but, not for any dark baryons!), may be used to
estimate $f_{\rm B}$.  From observations of the Sunyaev-Zeldovich
effect in X-ray clusters, \cite{grego} constrain the hot gas
fraction which \cite{skz} have used to estimate $f_{\rm B}$
and to derive a present-universe ($t_{0} \approx 10$~Gyr; 
$z \la 1$) baryon density.
\beq
\eta_{10}({\rm SNIa;Flat}) = 5.1 ^{+1.8} _{-1.4} ~~~ 
(\Omega_{\rm B}h^{2} = 0.019 ^{+0.007} _{-0.005}).
\eeq

\begin{figure}
  \centering
  \epsfysize=3.0truein
  \epsfbox{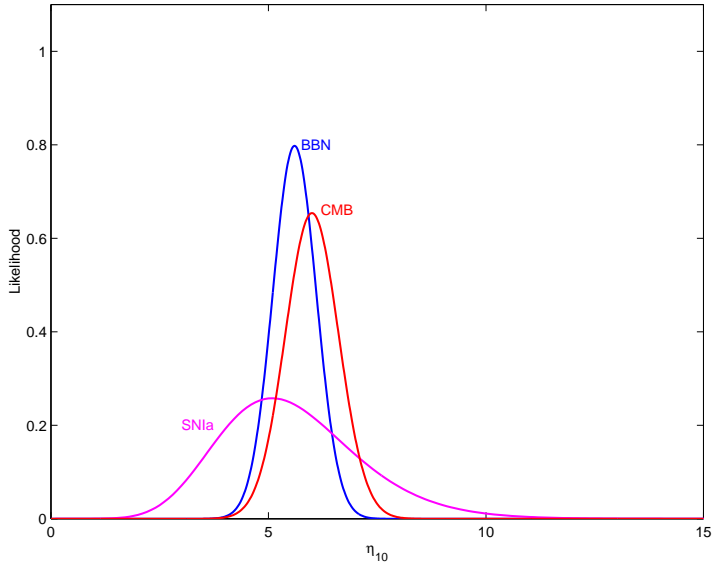}
  \caption{The likelihood distributions (normalized to equal areas 
under the curves) for the baryon-to-photon ratios derived from BBN, 
from the CMB, and for the present universe (SNIa) using the SNIa 
and X-ray cluster data, and the assumption of a flat universe.
}\label{fig:lik3}
\end{figure}

In Figure~\ref{fig:lik3} the corresponding likelihood distribution 
for the present universe baryon density is shown (``SNIa") along 
with those derived earlier from deuterium and BBN, and from the CMB 
fluctuation spectra.  Although the uncertainties are largest for this 
present-universe value, it is in excellent agreement with the other, 
independent estimates.

\section{Summary And Discussion}\label{sec:summ}

The abundances of the relic nuclides produced during BBN encode the 
baryon density during the first few minutes in the evolution of the 
universe.  Of these relics from the early universe, deuterium is the
baryometer of choice.  The deuterium abundance in relatively unprocessed
material, such as the high-z, low-Z QSO absorption line systems should
be very nearly primordial.  At present there are data for five such
systems.  Although the statistical accuracies of these data are high,
the dispersion among them in the derived D/H ratio is surprisingly
large, suggesting that systematic errors (interlopers?, complex velocity
structure?) may effect one or more of these determinations.  Nonetheless,
these data seem consistent with a primordial abundance (D/H)$_{\rm P} =
3.0 \pm 0.4 \times 10^{-5}$ (\cite[O'Meara \etal 2001]{O'M}) which was
adopted in \S\,\ref{sec:BBN} to derive $\eta_{10}({\rm BBN}) = 5.6 \pm 0.5$ 
($\Omega_{\rm B}h^{2} = 0.020 \pm 0.002$).  

Several hundred thousand years later, when the universe became transparent 
to the CMB radiation, the baryon density was imprinted on the temperature
fluctuation spectrum which has been observed by the COBE (\cite[Bennett 
\etal 1996]{cobe}), BOOMERANG (\cite[Netterfield \etal 2001]{BOOM}),
MAXIMA (\cite[Lee \etal 2001]{MAX}), and DASI (\cite[Halverson \etal
2001]{DASI}) experiments.  For determining the baryon density, these 
current CMB data have a precision approaching that of BBN: $\eta_{10}({\rm 
CMB}) = 6.0 \pm 0.6$ ($\Omega_{\rm B}h^{2} = 0.022 \pm 0.002$).  The excellent 
agreement between the BBN and CMB values (see Fig.~\ref{fig:lik2}) provides 
strong support for the standard model of cosmology.

In the present universe most baryons are dark ($\eta ({\rm LUM}) \ll 
\eta ({\rm BBN}) \approx \eta ({\rm CMB})$), so that estimates of the 
baryon density some 10 billion years after the expansion began are more 
uncertain and, often model-dependent.  In \S\,\ref{sec:sn1a} we combined 
an estimate of the total matter density ($\Omega_{\rm M}$) derived from 
the SNIa magnitude-redshift data (\cite[Perlmutter \etal 1997]{scp1}; 
\cite[Schmidt \etal 1998]{hiz}; \cite[Perlmutter \etal 1999]{scp2}) 
and the assumption of a flat universe ($\Omega_{\rm M} = 0.28 ^{+0.08} 
_{-0.07}$), with the universal baryon fraction inferred from X-ray 
observations of clusters of galaxies (\cite[Grego \etal 2001]{grego}) 
to derive $\eta_{10}({\rm SNIa;Flat}) = 5.1 ^{+1.8} _{-1.4}$ 
($\Omega_{\rm B}h^{2} = 0.019 ^{+0.007} _{-0.005}$) (\cite[Steigman, 
Kneller, \& Zentner 2000]{skz}).  Although of much lower statistical 
accuracy, this estimate of the present universe baryon density is in 
complete agreement with those from BBN and the CMB (see Fig.~\ref{fig:lik3}).
I note in passing that if the mass of dark baryons in clusters is
similar to the stellar mass, this present-universe baryon density
estimate would increase by $\sim 10\%$, bringing it into essentially
perfect overlap with the BBN and CMB values.
\begin{figure}
  \centering
  \epsfysize=3.5truein
  \epsfbox{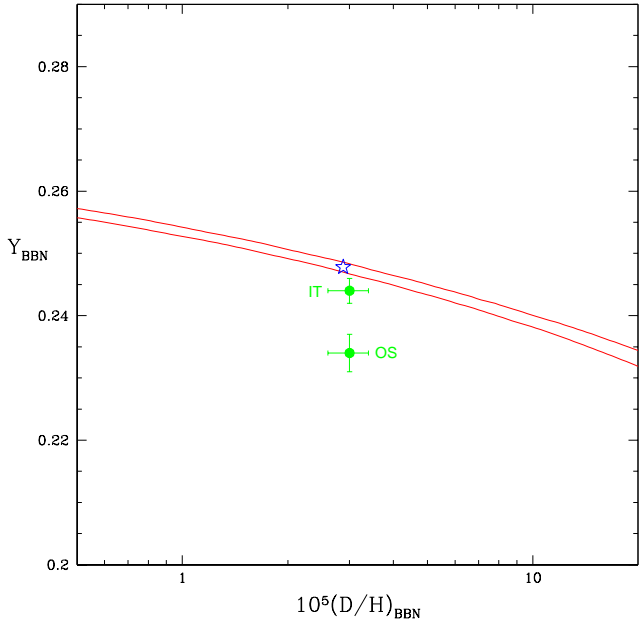}
  \caption{The BBN-predicted \4he mass fraction (Y$_{\rm BBN}$) 
  versus the deuterium abundance (by number with respect to hydrogen: 
  (D/H)$_{\rm BBN}$) is shown by the band from left-to-right.  The 
  star corresponds to Y$_{\rm BBN}$ and (D/H)$_{\rm BBN}$ for the 
  ``best" value of the universal density of baryons (see the text).  
  The data points are shown at the \cite{O'M} deuterium abundance 
  estimate and the IT and OS values for the helium abundance (see 
  the text).}\label{fig:hevsd}
\end{figure}

The concordance of the standard, hot, big bang cosmological model is 
revealed clearly by the overlapping likelihood distributions for the
universal density of baryons shown in Fig.~\ref{fig:lik3}.  As satisfying
as this agreement might be, it should impell us to action, not complacency.
How may we test further the standard model?  One way is to return to
the relic nuclides which, so far, have been set aside: \4he and \7li.

\begin{figure}
  \centering
  \epsfysize=3.5truein
  \epsfbox{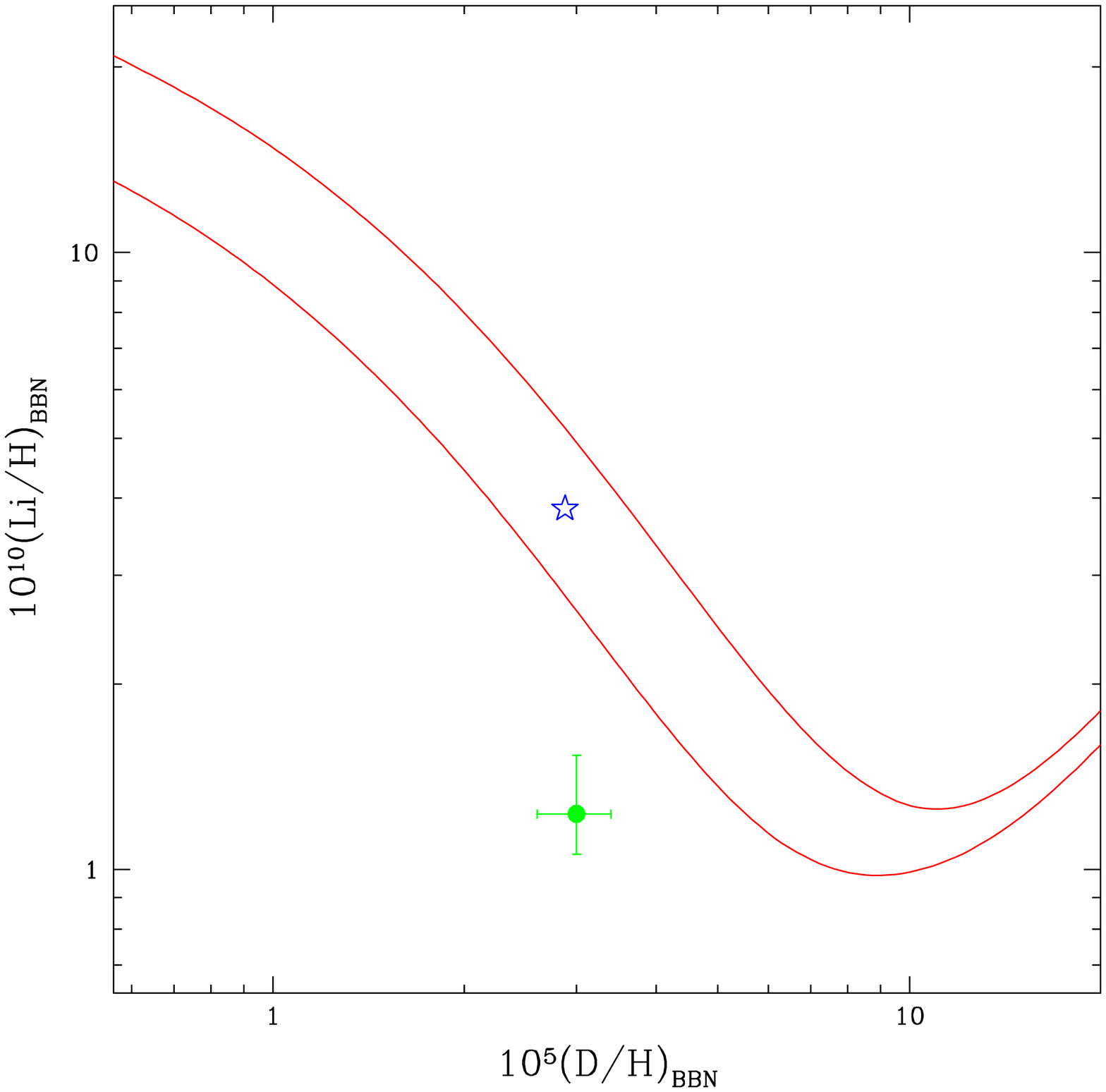}
  \caption{As in Fig.~\ref{fig:hevsd}, but for lithium versus deuterium.  
  The star is the BBN-predicted ``best" value for \7li and D, while the 
  \7li point is from \cite{ryan}.}\label{fig:livsd}
\end{figure}

Consistency among the three, independent baryon density estimates 
permits us to identify a ``best" value: $(\eta_{10})_{\rm best} = 5.7$ 
($(\Omega_{\rm B}h^{2})_{\rm best} = 0.021$).  For this baryon-to-photon 
ratio the BBN-predicted primordial abundances are: (D/H)$_{\rm P} = 2.9 
\times 10^{-5}$, Y$_{\rm P} = 0.248$, and (Li/H)$_{\rm P} = 3.8 \times 
10^{-10}$.  These ``best" estimates are shown by the ``stars" in Figures
~\ref{fig:hevsd} and \ref{fig:livsd}.  In those figures the bands reflect 
the BBN-predicted \4he vs. D and \7li vs D relations, including the nuclear 
and weak interaction physics uncertainties.  It is clear that the deuterium
abundance is in excellent agreement with current data.  What of \4he
and \7li?

At present there are two estimates for the primordial abundance of 
\4he based on large (nearly) independent data sets and analyses of 
low-metallicity, extragalactic \hii regions; see Fig.~\ref{fig:hevsd}.  
The ``IT" (\cite[Izotov, Thuan, \& Lipovetsky 1994]{itl}, \cite[Izotov 
\& Thuan 1998]{it}) estimate of Y$_{\rm P}({\rm IT}) = 0.244 \pm 0.002$ 
is only $2\sigma$ away from our best fit value of 0.248, while the ``OS" 
determination (\cite[Olive \& Steigman 1995]{os}, \cite[Olive, Skillman, 
\& Steigman 1997]{oss}, \cite[Fields \& Olive 1998]{fo}) of Y$_{\rm P}
({\rm OS}) = 0.234 \pm 0.003$ is nearly $5\sigma$ lower.  Although it may 
be tempting to dismiss the OS estimate, recent high quality observations 
of a relatively metal-rich \hii region in the SMC by \cite{ppr} reveal 
an abundance (Y$_{\rm SMC} = 0.2405 \pm 0.0018$) which is {\it lower} 
than the IT primordial value.   When this abundance is extrapolated to 
zero metallicity, \cite{ppr} find Y$_{\rm P}({\rm PPR}) = 0.2345 \pm 
0.0026$, in excellent -- albeit accidental! -- agreement with the OS 
value.  The comparisons among different observations and between theory 
and observations suggest that unaccounted systematic errors (underlying 
stellar absorption weakening the helium emission lines?) may have 
contaminated at least some of the data.

The comparison between theory and data for \7li poses much more of 
a challenge; see Fig.~\ref{fig:livsd}.  The data point plotted in
Fig.~\ref{fig:livsd}, from \cite{ryan}: (Li/H)$_{\rm P} = 1.23 ^{+0.68} 
_{-0.32} \times 10^{-10}$, is lower by a factor of three ($\sim 0.5$~dex)
than the standard model prediction (the ``star" in Fig.~\ref{fig:livsd}).
There are, however, good reasons to believe that the low value of the
lithium abundance derived by \cite{ryan} from observations of metal-poor,
{\it old} stars (\cite[Ryan, Norris \& Beers 1999]{rnb}) is not representative 
of the lithium abundance in the gas out of which these stars formed 
(\cite[Pinsonneault \etal 1999]{pwsn}; \cite[Salaris \& Weiss 2001]{weiss}; 
\cite[Pinsonneault \etal 2001]{pswn}; \cite[Theado \& Vauclair 2001]{vauclair}).
Using a model for lithium depletion via rotational mixing over the long
lifetimes of these stars, \cite{pswn} reanalyzed the \cite{rnb} data and
concluded that the primordial lithium abundance should be higher than the
\cite{ryan} value by $\approx 0.3 \pm 0.2$ dex, bringing the abundance
inferred from the earliest generation of stars in the Galaxy in closer
agreement with that expected for the ``best" estimate of the baryon 
density in the standard cosmological model.

\section{Conclusions}\label{sec:concl}

Increasingly precise observational data permit us to track the baryon
density from the big bang to the present.  At widely separated epochs
from the first few minutes, through the first few hundred thousand 
years, to the present universe, a consistent value emerges, accurate
to better than 10\%: $\eta_{10} \approx 5.7$ ($\Omega_{\rm B}h^{2}
\approx 0.021$).  Such a low baryon density ($\Omega_{\rm B} \approx 
0.04$) reinforces the need for non-baryonic dark matter ($\Omega_{\rm B} 
\la \Omega_{\rm M}/7$), which itself appears to be subdominant at present
to an unknown form of dark energy.  Precision cosmology has led us to an
extreme form of the Copernican Principle!   While this baryon density 
is fully consistent with present estimates of the primordial deuterium 
abundance, it challenges the data and ``astrophysics" used to determine 
the primordial abundances of \4he and \7li.  If unresolved by new data, 
or better astrophysics, these conflicts might be providing a peek at 
new physics beyond the standard models of cosmology and/or particle 
physics.  These are surely exciting times.

\begin{acknowledgments}

For valuable advice, assistance, and tutorials I wish to thank Gus Evrard, 
Jim Felten, Jim Kneller, Jeff Linsky, Joe Mohr, Paulo Molaro, Keith Olive, 
Manuel Peimbert, Max Pettini, Marc Pinsonneault, Bob Scherrer, Evan Skillman, 
David Tytler, Sueli Viegas, Terry Walker, and Andrew Zentner.  Praise 
is due Mario Livio and the efficient staff of the STScI for the smooth 
organization of an exciting meeting.  My research is supported by DOE 
grant DE-FG02-91ER-40690. 

\end{acknowledgments}


\begin{thebibliography}{}

\bibitem[Bennett \etal (1996)]{cobe}
\textsc{Bennett, C. L., \etal} 1996 ApJ {\bf 464}, L1.

\bibitem[Burles \& Tytler (1998a)]{bt98a}
\textsc{Burles, S. \& Tytler, D.} 1998a ApJ {\bf 499}, 699.

\bibitem[Burles \& Tytler (1998b)]{bt98b}
\textsc{Burles, S. \& Tytler, D.} 1998b ApJ {\bf 507}, 732.

\bibitem[Burles, Nollett, \& Turner (2001)]{BNT}
\textsc{Burles, S., Nollett, K. M., \& Turner, M. S.} 2001 Phys. Rev. 
{\bf D63}, 063512.

\bibitem[Cyburt, Fields, \& Olive (2001)]{cyburt}
\textsc{Cyburt, R. H., Fields, B. D., \& Olive, K. A.} 2001 preprint
(astro-ph/0102179).

\bibitem[D'Odorico, Dessauges-Zavadsky, \& Molaro (2001)]{DOd1} 
\textsc{D'Odorico, S., Dessauges-Zavadsky, M., \& Molaro, P.} 2001 
A\&A, {\bf 338}, L1.

\bibitem[Epstein, Lattimer, \& Schramm (1976)]{els}
\textsc{Epstein, R., Lattimer, J., \& Schramm, D. N.} 1976 Nature
{\bf 263}, 198.

\bibitem[Fields \& Olive (1998)]{fo}
\textsc{Fields, B. D. \& Olive, K. A.} 1996 ApJ {\bf 506}, 177.

\bibitem[Freedman \etal (2001)]{HST}
\textsc{Freedman, W. L., \etal} 2001 ApJ, {\bf 553}, 47.

\bibitem[Jedamzik \& Fuller (1997)]{jf}
\textsc{Jedamzik, K. \& Fuller, G.} 1997 ApJ {\bf 483}, 560.

\bibitem[Gloeckler \& Geiss (2000)]{gg}
\textsc{Gloeckler, G. \& Geiss, J.} 2000 Proceedings of IAU Symposium 
198, The Light Elements and Their Evolution (L. da Silva, M. Spite, and 
J. R. Medeiros eds.; ASP Conference Series), p. 224.

\bibitem[Grego \etal (2001)]{grego}
\textsc{Grego, L., \etal} 2001 ApJ {\bf 552}, 2.

\bibitem[Halverson \etal (2001)]{DASI}
\textsc{Halverson, N. W., \etal} 2001 preprint (astro-ph/0103305).

\bibitem[Izotov, Thuan, \& Lipovetsky (1994)]{itl}
\textsc{Izotov, Y. I., Thuan, T. X., \& Lipovetsky, V. A.} 1994 
ApJ {\bf 435}, 647.

\bibitem[Izotov \& Thuan (1998)]{it}
\textsc{Izotov, Y. I. \& Thuan, T. X.} 1998 ApJ {\bf 500}, 188.

\bibitem[Kirkman \etal (2001)]{antihid}
\textsc{Kirkman, D., \etal} 2001 preprint (astro-ph/0104489).

\bibitem[Kneller, Scherrer, Steigman, \& Walker (2001)]{kssw}
\textsc{Kneller, J. P., Scherrer, R. J., Steigman, G., \& Walker, 
T. P.} 2001 preprint (astro-ph/0101386).
 
\bibitem[Lee \etal (2001)]{MAX}
\textsc{Lee, A. T., \etal} 2001 preprint (astro-ph/0104459).
 
\bibitem[Levshakov, Dessauges-Zavadsky, D'Odorico, \& Molaro (2001)]{DOd2}
\textsc{Levshakov, S. A., Dessauges-Zavadsky, M., D'Odorico, S., 
\& Molaro, P.} 2001 preprint (astro-ph/0105529). 

\bibitem[Linsky \& Wood (2000)]{linsky}
\textsc{Linsky, J. L. \& Wood, B. E.} 2000 Proceedings of IAU Symposium 
198, The Light Elements and Their Evolution (L. da Silva, M. Spite, and 
J. R. Medeiros eds.; ASP Conference Series), p. 141.

\bibitem[Netterfield \etal (2001)]{BOOM}
\textsc{Netterfield, C. B., \etal} 2001 preprint (astro-ph/0104460).

\bibitem[Olive \& Steigman (1995)]{os}
\textsc{Olive, K. A. \& Steigman, G.} 1995 ApJ Suppl. {\bf 97}, 49.

\bibitem[Olive, Skillman, \& Steigman (1997)]{oss}
\textsc{Olive, K. A., Skillman, E., \& Steigman, G.} 1997 ApJ {\bf 483}, 
788.

\bibitem[Olive, Steigman, \& Walker (2000)]{OSW}
\textsc{Olive, K. A., Steigman, G., \& Walker, T. P.} 2000 Phys.
Rep. {\bf 333}. 389.

\bibitem[O'Meara \etal (2001)]{O'M}
\textsc{O'Meara, J. M., \etal} 2001 ApJ {\bf 552}, 718.

\bibitem[Peimbert, Peimbert, \& Ruiz (2000)]{ppr}
\textsc{Peimbert, M., Peimbert, A., \& Ruiz, M. T.} 2000 ApJ 
{\bf 541}, 688.

\bibitem[Perlmutter \etal (1997)]{scp1}
\textsc{Perlmutter, S., \etal} 1997 ApJ {\bf 483}, 565.

\bibitem[Perlmutter \etal (1999)]{scp2}
\textsc{Perlmutter, S., \etal} 1999 ApJ {\bf 517}, 565.

\bibitem[Persic \& Salucci (1992)]{ps}
\textsc{Persic, M. \& Salucci, P.} 1992 MNRAS {\bf 258}, 14P.

\bibitem[Pettini \& Bowen (2001)]{PB}
\textsc{Pettini, M. \& Bowen, D. V.} 2001 preprint (astro-ph/0104474).

\bibitem[Pinsonneault \etal (1999)]{pwsn}
\textsc{Pinsonneault, M. H., Walker, T. P., Steigman, G., \&
Narayanan, V. K.} 1999 ApJ {\bf 527}, 180.

\bibitem[Pinsonneault \etal (2001)]{pswn}
\textsc{Pinsonneault, M. H., Steigman, G., Walker, T. P., \&
Narayanan, V. K.} 2001 preprint (astro-ph/0105439).

\bibitem[Ryan, Norris, \& Beers (1999)]{rnb}
\textsc{Ryan, S. G., Norris, J. E., \& Beers, T. C.} 1999 ApJ
{\bf 523}, 654.

\bibitem[Ryan \etal (2000)]{ryan}
\textsc{Ryan, S. G., Beers, T. C., Olive, K. A., Fields, B. D., 
\& Norris, J. E.} 2000 ApJ {\bf 530}, L57.

\bibitem[Salaris \& Weiss (2001)]{weiss}
\textsc{Salarais, M. \& Weiss, A.} 2001 preprint (astro-ph/0104406).

\bibitem[Sanders (2001)]{sanders}
\textsc{Sanders, R.H.} 2001 preprint (astro-ph/0106558). 

\bibitem[Schmidt \etal (1998)]{hiz}
\textsc{Schmidt, B. P., \etal} 1998 ApJ {\bf 507}, 46.

\bibitem[Steigman \& Felten (1995)]{sf}
\textsc{Steigman, G. \& Felten, J. E.} 1995 Spa. Sci. Rev. {\bf 74}, 245.

\bibitem[Steigman, Hata, \& Felten (1999)]{shf}
\textsc{Steigman, G.,  Hata, N., \& Felten, J. E.} 1999 ApJ {\bf 510}, 564.

\bibitem[Steigman, Kneller, \& Zentner (2001)]{skz}
\textsc{Steigman, G., Kneller, J. P., \& Zentner, A.} 2001 preprint 
(astro-ph/0102152).

\bibitem[Steigman, Walker, \& Zentner (2000)]{swz}
\textsc{Steigman, G., Walker, T. P., \& Zentner, A.} 2000 preprint 
(astro-ph/0012149).

\bibitem[Theado \& Vauclair (2001)]{vauclair}
\textsc{Theado, S. \& Vauclair, S.} 2001 preprint (astro-ph/0106080).

\bibitem[Webb \etal (1997)]{hid}
\textsc{Webb, J. K., Carswell, R. F., Lanzetta, K. M., Ferlet, R.,
Lemoine, M., Vidal-Madjar, A., \& Bowen, D. V.} 1997 Nature {\bf 388}, 
250.

\bibitem[Weinberg, Miralda-Escud\'e, Hernquist, \& Katz (1997)]{wein}
\textsc{Weinberg, D. H., Miralda-Escud\'e, J., Hernquist, L., \& Katz, 
N.} 1997 ApJ {\bf 490}, 564.


\end{thebibliography}
\end{document}